\documentclass[aps,prl,twocolumn,showpacs,showkeys]{revtex4}

\usepackage{amssymb}
\usepackage{times}
\usepackage{mathptmx} 

\newcommand{\smfrac}[2]{\mbox{$\frac{#1}{#2}$}}

\newcommand{\ket}[1]{|#1\rangle}

\newcommand{\Prob}{\mathrm{Prob}}
\newcommand{\IP}{\textsc{IP}}
\newcommand{\Eq}{\textsc{Eq}}
\newcommand{\HAT}[1]{#1}

\begin{document}
\title{Implausible Consequences of Superstrong Nonlocality}
\author{Wim van Dam}
\email[Email: ]{vandam@cs.ucsb.edu}
\affiliation{Department of Computer Science, University of California at 
Santa Barbara, Santa Barbara, CA 93106-5110, USA}

\begin{abstract}
This Letter looks  at the consequences of 
so-called `superstrong nonlocal correlations', which are 
hypothetical violations of Bell/CHSH inequalities that are stronger 
than quantum mechanics allows, 
yet weak enough to prohibit faster-than-light communication.  
It is shown that the existence of maximally superstrong correlated 
bits implies that all distributed computations can be 
performed with a trivial amount of communication, i.e.\ with one bit.
If one believes that Nature does not allow such a computational 
`free lunch', then the result in the Letter gives a reason why
superstrong correlation are indeed not possible.
\end{abstract}

\pacs{03.65.Ud, 03.65.Ta, 03.67.Hk, 03.67.Mn}
\keywords{foundations of quantum mechanics, nonlocality, 
communication complexity}

\maketitle

The Clauser-Horne-Shimony-Holt (CHSH) inequality \cite{CHSH} for classical
theories gives the following upper bound on the strength of correlations 
between two space-like separated experiments, which can be violated 
by quantum mechanics.
Imagine two parties Alice and Bob ($A$ and $B$) 
that share a distributed system 
$\Phi_{AB}$. 
Each party can independently perform one out of two measurements 
on their part of the system, such that in 
total there are four experimental set-ups that can apply to the
combined system: $(m^A_0,m^B_0)$, $(m^A_0,m^B_1)$, $(m^A_1,m^B_0)$ 
and $(m^A_1,m^B_1)$. For each measurement on each side 
there are two possible outcomes, which are labeled ``$0$'' and ``$1$''. 
The parties repeat the experiment many times using the different settings, 
thus obtaining an accurate estimation of all the possible
correlations between the different measurements and their outcomes. 
As it is understood that for each trial $A$ and $B$ always use the same 
state-preparation of $\Phi_{AB}$, the conditional part will be omitted
when expressing the probabilities of the various outcomes.  
Hence, the probability that both Alice and Bob measure a ``one''
when they use the measurement settings $m^A_0$ and $m^B_1$  is 
denoted simply by $\Prob({\HAT{m}}^A_0=1,{\HAT{m}}^B_1 = 1)$. 

The main result of Bell \cite{Bell} and CHSH \cite{CHSH} is that {for any 
local, hidden variable theory about $\Phi_{AB}$ and the measurements 
$m^A$ and $m^B$, the following inequality must hold:}
\begin{eqnarray} \label{eq:local}
\sum_{x,y\in\{0,1\}}\Prob({\HAT{m}}^A_x + {\HAT{m}}^B_y \equiv x\cdot y)&\leq& 3,
\end{eqnarray} 
where we interpret the binary values 
as elements of `modulo 2 calculations' such that $1+1\equiv 0$.
Quantum mechanics allows a violation of the bound of Equation~\ref{eq:local} by
\begin{eqnarray*}
\sum_{x,y\in\{0,1\}}{\Prob({\HAT{m}}^A_x + {\HAT{m}}^B_y \equiv x\cdot y)} 
& = &  2 + \sqrt{2} ~\approx~ 3.41,
\end{eqnarray*}
if $A$ and $B$ use, for example, the entangled pair of quantum bits
$\ket{\Phi_{AB}}=\smfrac{1}{\sqrt{2}}(\ket{00}+\ket{11})$
and a suitable set of measurement projectors $m$.  
Besides the fact that this result proves that the theory of 
quantum mechanics cannot be phrased as a local theory, the more 
important conclusion is that the nonlocality of Nature can be verified 
experimentally (as has been done many 
times \cite{ADR:etobiutva,FC:etolhvt}).
This experimental aspect is the more relevant side of the matter as it is 
not inconceivable that in the future we will have to replace the theory of 
quantum mechanics by a more accurate or more general model of Nature, 
making the nonlocality of quantum mechanics irrelevant.
But no matter its exact formulation, the succeeding theory will have to 
agree with our experimental results; 
and as the empirical data by itself rules out a local explanation, 
any proper future candidate theory will have to be nonlocal as well.
From this perspective, which we could call 
`nonlocality-without quantum physics',
we should consider all possible violations of 
Equation~\ref{eq:local}, not just the 
``$2+\sqrt{2}\not\leq 3$'' violation
of quantum mechanics.
In this Letter we look at the plausibility of 
\emph{superstrong nonlocality} where the nonlocal correlations 
are stronger than those allowed by the theory of quantum physics.

In a series of articles \cite{PR:qnaaa,PR:trea,RP:naaafqt}, 
Sandu Popescu and Daniel Rohrlich ask
the question why Nature seems to allow a violation of the CHSH 
inequality with a correlation term of $2+\sqrt{2}$, but not with 
more.
(See the article by Boris Cirel'son \cite{C:qgobi} for a 
proof that $2+\sqrt{2}$ is indeed the quantum mechanical limit.)
They ask themselves \cite{PR:trea}: ``\dots Could the requirement 
of relativistic causality restrict the violation to [$2+ \sqrt{2}$] 
instead of $4$?'' 
 Such a result would be great step towards a better understanding of
Nature for ``\dots If so, then nonlocality and causality would together 
determine the quantum violation of the CHSH inequality, and we would 
be closer to a proof that they determine all of quantum mechanics.''
Perhaps surprisingly, this turns out not to be the case. 
The authors prove this by constructing a toy-theory where the
nonlocality Inequality~\ref{eq:local} is surpassed by a correlation 
value of $4$.
The non-zero probabilities of this super-nonlocal theory are simply
\begin{eqnarray} \label{eq:sstrong}
\left.{\begin{array}{l}
\Prob({\HAT{m}}^A_x=0,{\HAT{m}}^B_y = 0) = \frac{1}{2} \\
\Prob({\HAT{m}}^A_x=1,{\HAT{m}}^B_y=1) = \frac{1}{2} 
\end{array}}\right\}
& &\textrm{if~}xy\in\{00,01,10\}, \nonumber \\
& & \nonumber \\
\left.{\begin{array}{l}
\Prob({\HAT{m}}^A_1=0,{\HAT{m}}^B_1 = 1) = \frac{1}{2} \\
\Prob({\HAT{m}}^A_1=1,{\HAT{m}}^B_1=0) = \frac{1}{2} 
\end{array}}\right\} & & 
 \textrm{if~}xy=11.
\end{eqnarray}
This leads indeed to the maximally violating correlation value
\begin{eqnarray} \label{eq:sstrong2}
\sum_{x,y\in\{0,1\}}{\Prob({\HAT{m}}^A_x + {\HAT{m}}^B_y \equiv x\cdot y)} 
& = &  4,
\end{eqnarray}
while the randomization of the outcomes still prevents Alice or Bob 
from transferring information to the other party without the use of 
conventional communication.  In fact, the probability distribution of 
Equation~\ref{eq:sstrong} is the only possible solution 
if we want to combine a correlation value of $4$ with the 
preservation of causality.

So, if causality is still respected with the superstrong correlations 
of Equation~\ref{eq:sstrong}, why does Nature not allow it? Are there any
obvious first principles that forbid a violation stronger than 
that of quantum mechanics?  
When trying to answer this question in a meaningful way, it is important 
to remember to ignore everything one knows about  quantum mechanics.  
As explained above, the point 
is to consider all possible (future) physical theories, not just 
the contemporary one.  Cirel'son's bound already shows us that quantum 
mechanics is incompatible with a violation of Equation~\ref{eq:local}
that goes beyond ``$2+\sqrt{2}\not\leq 3$'', hence it is not interesting 
to derive a contradiction under assumptions that use features of quantum 
mechanics like the superposition principle, linearity, et cetera. 
Instead, we want to assume nothing else but a violation 
of Equation~\ref{eq:local} by a value greater than $2+\sqrt{2}$.

In this Letter we look at the consequences of superstrong nonlocality 
for the theory of \emph{communication complexity,} which describes how 
much communication is needed to evaluate a distributed function $f$.
More specifically, consider a Boolean function 
$f:\{0,1\}^n\times\{0,1\}^n\rightarrow \{0,1\}$, which has as 
input two $n$-bit strings $\vec{x},\vec{y}\in\{0,1\}^n$.
If $A$ possesses the $x$-string and $B$ the $y$-string, how many
bits do $A$ and $B$ have to exchange in order to determined 
the function value $f(\vec{x},\vec{y})$?  
How to answer this question---which depends on the specific 
function and the resources of $A$ and $B$---is studied 
in the field of (quantum) communication complexity \cite{KN:cc}.  
For certain $f$ it has been shown that quantum entanglement can 
reduce the amount of classical information that $A$ and 
$B$ need to exchange to evaluate $f$ \cite{CB}, while for other functions 
the quantum complexity is effectively the same as the 
classical complexity.  An example of latter is the 
Inner Product function $\IP_n:\{0,1\}^n\times\{0,1\}^n\rightarrow \{0,1\}$, 
which is defined by 
\begin{eqnarray*}
\IP_n(x_1\cdots x_n,y_1\cdots y_n) 
& \equiv & \sum_{i=1}^n{x_i\cdot y_i}.
\end{eqnarray*}
Even if we allow $A$ and $B$ to use an unlimited amount of 
entangled qubits, the communication complexity will still be 
$n$ bits \cite{IP} (which is the maximum possible complexity as 
$B$ can always send $A$ all of the $n$ bits of his input $\vec{y}$, 
after which $A$ evaluates $f(\vec{x},\vec{y})$ on her side).

Here it will be shown that a maximum violation of the CHSH 
Inequality~\ref{eq:local} (according to the ``$4\not\leq 3$''
of Equation~\ref{eq:sstrong2})
leads to a situation where the notion of 
communication complexity is vacuous: all distributed 
decision problems can be solved with $100\%$ accuracy with 
only one bit of communication. (Note that at least one bit needs
to be communicated if we want to preserve causality.)
To prove our result, we first describe a way of expressing all 
possible distributed functions in a standard format that coincides 
with the inner product problem for two parties. 
Then we will see how, with superstrong 
correlations, the $\IP$ problem (and hence all problems) can be solved 
with the minimal amount of one bit of communication from Bob to Alice. 
The results in this Letter were mentioned earlier in the Ph.D.\ thesis
of the author \cite{PhD}.

Any Boolean function $f:\{0,1\}^n\times\{0,1\}^n\rightarrow\{0,1\}$
can be expressed as a multi-variable polynomial with modulo two
arithmetic (where $1+1=2\equiv 0$). This is most easily seen by 
the fact that elementary Boolean operations like \textsc{and},
\textsc{or}, \textsc{not} or `equivalence' can be calculated 
with addition and  multiplication over $\{0,1\}$:
\begin{eqnarray*}
& & \left\{{
\begin{array}{rclcrcl}
(x \textsc{ and } y)   &\equiv &  x\cdot y, &  &
(x \textsc{ or } y)  & \equiv & x + y + x\cdot y, \\ 
& & \vspace{-10pt}\\
\textsc{not}(x)  &\equiv & 1+x, & & 
(x \Leftrightarrow y) & \equiv & 1+x+y,
\end{array}}\right.
\end{eqnarray*}
where the value $1$ means ``True'' and $0$ means ``False''.
Just as any Boolean function $f:\{0,1\}^n \rightarrow \{0,1\}$ can be 
constructed from those primitives,
so can $f$  be constructed from the elementary 
$\bmod{2}$ operations ``$+$'' and ``$\cdot$''.
The $2$-bit equivalence relation \Eq, for example, thus
becomes
\begin{eqnarray*}
\Eq(x_1x_2,y_1y_2) & = & 
(x_1 \Leftrightarrow y_1) \textsc{and} (x_2\Leftrightarrow y_2)
\\
&\equiv&  (1+x_1+y_1)\cdot(1+x_2+y_2).
\end{eqnarray*}
Furthermore, we can rewrite such polynomials as a finite summation 
of products $f(\vec{x},\vec{y}) \equiv 
\sum_i P_i(\vec{x})\cdot Q_i(\vec{y})$, where  $P_i$ are polynomials
in $\vec{x}\in\{0,1\}^n$ and $Q_i$ are monomials in $\vec{y}\in\{0,1\}^n$.  
In total there are $2^n$ different monomials
$Q_i(\vec{y}) = \prod_{j\in S}{y_j}$ 
that we have to consider (one for each subset $S\subseteq\{1,\dots,n\}$),
hence  the index $i$ in the summation can be limited to 
$1\leq i \leq 2^n$. This gives us a way of 
representing the function $f$ as an inner product problem of input size
$2^n$:
\begin{eqnarray} \label{eq:ipred}
f(x_1\cdots x_n,y_1\cdots y_n) & \equiv & 
\sum_{i=1}^{2^n}{P_i(\vec{x})\cdot Q_i(\vec{y})},
\end{eqnarray}
with $\vec{x},\vec{y}\in\{0,1\}^n$.
For the example of the $2$-bit equality function $\Eq$ this is shown by
\begin{eqnarray*}
\Eq(x_1x_2,y_1y_2) & \equiv & 
(1+x_1+y_1)\cdot (1+x_2+y_2) \\
& \equiv & 
(1+x_1+x_2+x_1x_2)\cdot 1 + 1\cdot y_1y_2\\
& & + (1+x_2)\cdot y_1 + (1+x_1)\cdot y_2 \\
& \equiv &
\sum_{i=1}^4{P_i(x_1,x_2)\cdot Q_i(y_1,y_2)}.
\end{eqnarray*}
We can view this as an inner product problem because all the 
bit values $P_i(\vec{x})$ are known to Alice 
and all the values $Q_i(\vec{y})$ are known to Bob 
without the need for any communication between them.
Hence, if $A$ and $B$ are able to compute the $\IP$ function for 
input sizes of $2^n$ with one bit of communication, then they are 
also  able to calculate any decision problem
$f:\{0,1\}^n\times\{0,1\}^n \rightarrow \{0,1\}$ 
with a single bit of information exchange.
Next we will see that this indeed possible 
with a maximum violation of the CHSH inequality. 

Assume a model of Nature where the probabilities of
Equations~\ref{eq:sstrong} and \ref{eq:sstrong2}
are applicable, and hence where the correlation
\begin{eqnarray*}
\Prob({\HAT{m}}^A_x + {\HAT{m}}^B_y \equiv x\cdot y) & = & 1
\end{eqnarray*}
holds for all $x,y\in\{0,1\}$.
In such a world Alice and Bob (with input bits ${x}$ and ${y}$) 
can perform two separated  measurements on their 
super-correlated states that yield the outcomes 
$\alpha$ and $\beta$ obeying $\alpha+ \beta \equiv x\cdot y$. 
From this it follows that in the case of the inner product function 
$\IP_N$ on strings of length $N$, 
Alice and Bob can
perform $N$ measurements on $N$ super-correlated particle pairs 
in order to obtain---without any communication---a collection of bit values 
$\alpha_i$ and $\beta_i$, with $\alpha_i + \beta_i \equiv x_i\cdot y_i$
for every $1\leq i\leq N$.
The commutativity of addition (modulo two) allows the 
following regrouping of the bits by the two separated sides of the communication
protocol: 
\begin{eqnarray*}
\IP_N(x_1\cdots x_N,y_1\cdots y_N) & \equiv & 
\sum_{i=1}^{N}{x_i\cdot y_i}  \\ 
& \equiv &
\sum_{i=1}^{N}{(\alpha_i + \beta_i)} \\
&\equiv &
\underbrace{\left(\sum_{i=1}^{N}{\alpha_i}\right)}_{\textrm{Alice's~side}} + 
\underbrace{\left(\sum_{i=1}^{N}{\beta_i}\right)}_{\textrm{Bob's~side}}.
\end{eqnarray*}
Because Bob can construct and add his $\beta_i$ values without requiring 
any information from Alice, he can therefore compute the value 
$b \equiv \sum_i{\beta_i}$ by himself and broadcast this single 
bit to Alice.
She, on her part, creates the $\alpha_i$ values and finishes the 
protocol with the errorless conclusion 
$\IP(x,y) \equiv b + \sum_i{\alpha_i}$.

We just saw how the $\IP$ function has a communication complexity 
of one bit for every finite input size $N$ in the setting of superstrong
correlations. 
Hence, we can apply the reduction shown earlier
to reach the result that any distributed decision problem $f(\vec{x},\vec{y})$ 
can be exactly computed with a single bit of communication.
Equation~\ref{eq:ipred} tells us that we can rewrite the function $f$ to
$f(\vec{x},\vec{y}) \equiv  
\sum_i{P_i(\vec{x})\cdot Q_i(\vec{y})}$.
As Bob can compute all the $Q_i$ values by himself, he and Alice 
can also remotely and independently create the $\alpha_i$ and $\beta_i$ 
values such that $\alpha_i + \beta_i \equiv P_i(x)\cdot Q_i(y)$ for all 
$1\leq i \leq 2^n$. After the appropriate regrouping of the sum, 
 Equation~\ref{eq:ipred} then becomes
\begin{eqnarray*}
f(x_1\cdots x_n, y_1\cdots y_n) & \equiv & 
\underbrace{\left(\sum_{i=1}^{2^n}{\alpha_i}\right)}_{\textrm{Alice's~side}} 
 + 
\underbrace{\left(\sum_{i=1}^{2^n}{\beta_i}\right)}_{\textrm{Bob's~side}}.
\end{eqnarray*}
It should now be clear that Bob can compute the bit 
$b \equiv \sum_i{\beta_i}$ by himself and then communicate it to Alice
who, just as for the $\IP$ function, concludes with 
$f(\vec{x},\vec{y})\equiv b + \sum_i \alpha_i$. 

This finishes the proof that
with the help of the superstrong correlations of 
Equation~\ref{eq:sstrong} any distributed function can be decided
on Alice's side without error after only one bit of communication 
from Bob. It is true that in this protocol the amount of resources 
(the super-correlated states) can grow exponentially with the input 
size $n$ but this does not effect the conclusion that the 
communication complexity---after the inputs are distributed---is 
minimal. 

We can now rephrase our original question as:
why would Nature not allow super-efficient 
distributed computing?
It is not clear if there is a convincing answer
to this, as it does not seem to conflict with 
any physical intuition. 
However, trivial communication complexity 
does prohibit the existence of an intrinsic 
`complexity' for distributed tasks. 
Even though we need an exponential amount of 
prior superstrong nonlocality (as is indeed sometimes 
the case), the solution of all possible distributed 
functions with a single bit of communication does 
contradict our experiences that certain computational
tasks are harder than other ones.
Similar as in computability theory, there is a hierarchy 
of different complexity classes of communication problems \cite{BFS:ccicct}.
Such hierarchies are at the core of theoretical computer science,
and their absence---as happened here by assuming
superstrong correlations---goes squarely against the worldview
and experience of probably all researchers in the field of
complexity theory. 
 
If we accept the absence of intrinsic complexity as 
an argument against superstrong correlations, then 
it is natural to wonder if we can obtain similar results 
for correlations that are less strong than those of 
Equation~\ref{eq:sstrong2}.
It is tempting to speculate that a more detailed analysis 
would reveal that the $2+\sqrt{2}$ of quantum mechanics is a critical value
that separates trivial from nontrivial communication complexity,
which would give an argument for Cirel'son's nonlocality 
bound  without referring to quantum mechanics.
Recent work by Buhrman et al.\ \cite{Buhrman} shows that trivial communication
complexity can indeed be achieved with a correlation value 
strictly less than $4$. 
Because there is a tight connection between the communication complexity
of distributed functions and the depth of circuits for 
these problems \cite{KN:cc} it also possible to consider the 
implications of superstrong nonlocality for computational complexity.

\end{document}